\documentclass[reprint,aip,pre]{revtex4-1}

\usepackage{bm}
\usepackage[]{graphicx}
\usepackage{amsmath}
\usepackage{amssymb}
\usepackage{color}
\usepackage{soul}

\begin{document}

\title{Chemotactic Maneuverability of Sperm}

\author{Jeffrey S. Guasto}
\affiliation{Department of Civil \& Environmental Engineering, Massachusetts Institute of Technology,\\ Cambridge, Massachusetts 02139, USA}
\author{Jeffrey A. Riffell}
\affiliation{Department of Biology, University of Washington,\\ Seattle, Washington 98195, USA}
\author{Richard K. Zimmer}
\affiliation{\mbox{Department of Ecology \& Evolutionary Biology, University of California, Los Angeles,}\\ Los Angeles, California 90095, USA}
\author{Roman Stocker}
\affiliation{Department of Civil \& Environmental Engineering, Massachusetts Institute of Technology,\\ Cambridge, Massachusetts 02139, USA}

\begin{abstract}
In this fluid mechanics video, we explore the kinematics of chemotaxing sperm cells (sea urchin, \textit{Arbacia punctulata}) swimming in a chemoattractant gradient. 
We demonstrate that the complex swimming trajectories resulting in chemotactic behavior (`turn-and-run' motility) are comprised of several distinct flagellar maneuvers.
These motility patterns likely play an important role optimizing chemotaxic motility and navigation, when the sperm cells are subjected external fluid flows.
\end{abstract}

\maketitle

Fertilization is one of the most important, yet least understood biological process.
It plays a crucial role in ecosystem dynamics and human reproduction.
In many marine invertebrates (external fertilizers) as well as mammals (e.g. humans), motile sperm navigate toward eggs via chemical factor gradients emitted by the eggs, a process called chemotaxis.
While this behavior is known in many species, the nature of the active maneuverability exhibited by swimming sperm is not well understood.
In this study, we demonstrate that sea urchin sperm (\textit{Arbacia punctulata}) exhibit several distinct swimming patterns, which when used in concert, allow them to chemotax up chemical gradients.

\begin{figure}[h]
\centering
\includegraphics[keepaspectratio,width=8.5cm]{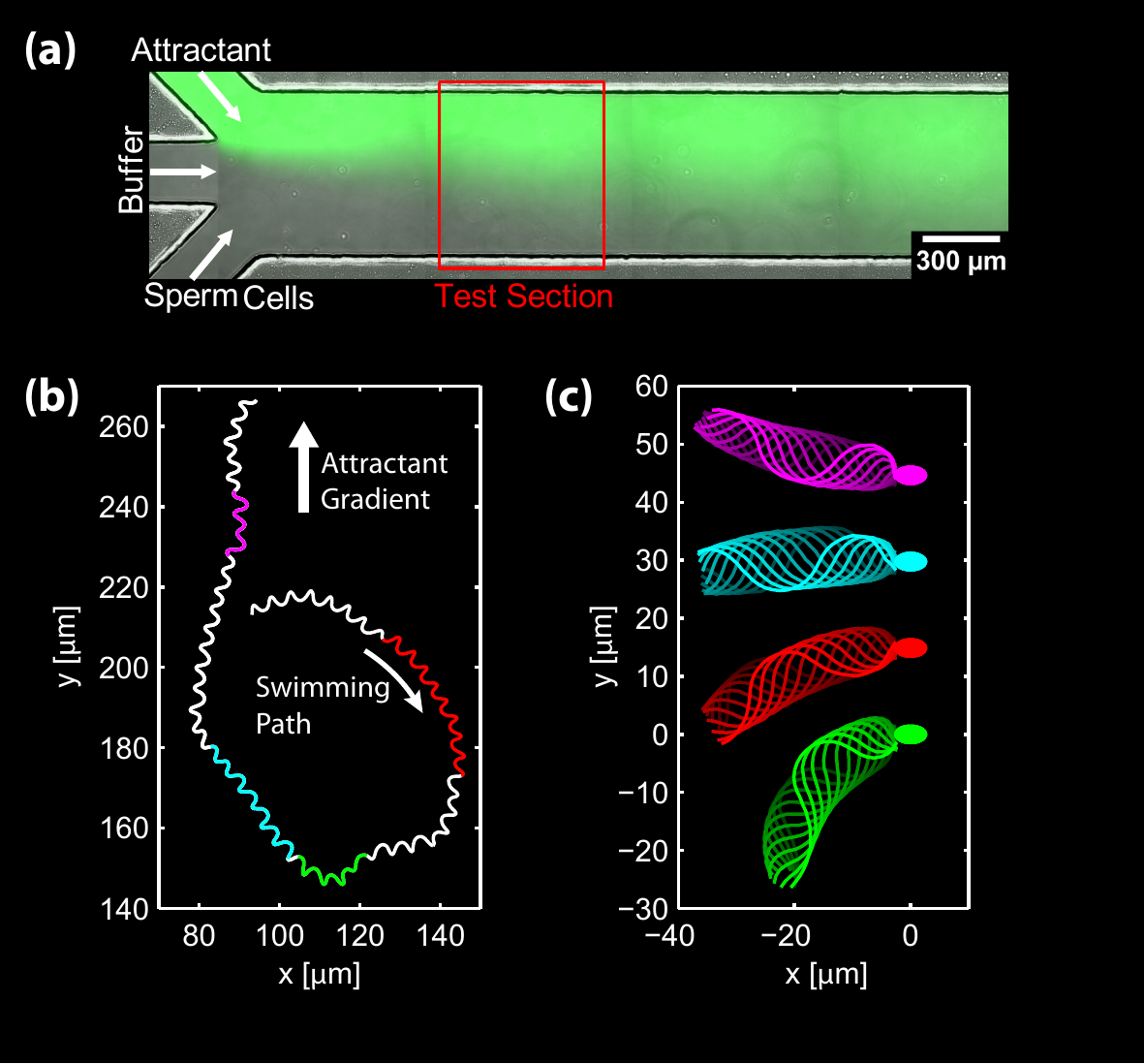}
\caption{(color) Cell motility and flagellar kinematics during chemotactic sperm swimming. 
(a) Sea urchin sperm (\textit{A. punctulata}) are exposed to precisely-controlled chemottractant gradients (resact) created using a three-inlet PDMS microchannel. 
(b) Sperm chemotactic motility relative to the attractant gradient is observed using high-speed imaging to capture flagellar kinematics. 
(c) Several distinct maneuvers are exhibited by the sperm including: sharp turns (green), arcs (red), straight runs (cyan), and opposite-handed turns (magenta). 
Flagellar waveforms are shown in the cell swimming frame over one beat cycle.} 
\label{Fig}
\end{figure}

Precisely-controlled gradients of a known chemoattractant (resact) are generated via a three-inlet polydimethylsiloxane (PDMS) microchannel (600 $\mu$m wide $\times$ 100 $\mu$m high test section). 
Filtered sea water (FSW, 0.2 $\mu$m), resact ($10^{-8}$ M in FSW), and sperm (diluted $10^4$ $\times$ in FSW) are flowed at equal rates (0.03 $\mu$L/min) into the test section of the channel [Fig. \ref{Fig}(a)].
Sperm are observed on the lower surface of the channel using phase-contrast microscopy ($40\times$, 0.6 NA).
Cell bodies are $\approx 2$ $\mu$m wide $\times$ 4 $\mu$m long, and the flagella ($\approx 50$ $\mu$m long, $\approx 250$ nm diameter) beat at 50-60 Hz propelling the cells at $\approx 250$ $\mu$m/s.
The bodies and flagella are imaged at high speed (750 fps) to capture the flagellar beating, and tracked with high accuracy using quantitative image analysis and particle tracking methods.

In the absence of any chemoattractant, the sperm robustly swim in 50-100 $\mu$m circular trajectories at the substrate surface. 
This behavior is thought to be used for sensing spatiotemporal attractant gradients \cite{Bohmer2005}. 
When exposed to an attractant gradient cells continue this circular swimming motion, but bias their trajectories with long runs, when oriented in the gradient direction (`turn-and-run' motility) [Fig. \ref{Fig}(b)]. 
During this chemotactic migration, at least four distinct trajectory components are observed: (\textit{i}) arcs, (\textit{ii}) sharp turns, (\textit{iii}) straight runs, and (\textit{iv}) opposite-handed turns. 
The observed mean flagellar curvature [Fig. \ref{Fig}(c)] is qualitatively positively correlated with the swimming trajectory curvature for each different behavior.
The exact nature of these chemotactic maneuvers likely plays an important role in sperm chemotaxis in fluid flows (i.e. shear), which is known to dramatically affect a sperm cell's ability to fertilize an egg \cite{Riffell2007}.
This work was supported by the NSF.

%\vspace{-0.11in}


\begin{thebibliography}{2}

\bibitem{Bohmer2005} B\"{o}hmer \textit{et al.}, ``Ca$^{2+}$ Spikes in the Flagellum Control Chemotactic Behavior of Sperm,'' EMBO J. {\bf 24}, 2741-2752 (2005).

\bibitem{Riffell2007} J. A. Riffell and R. K. Zimmer, ``Sex and Flow: the Consequences of Fluid Shear for Sperm-Egg Interactions,'' J. Exp. Bio. {\bf 210}, 3644-3660 (2007).

\end{thebibliography}
\end{document}